\documentclass[conference,10pt]{IEEEtran}
\usepackage{epsfig,rotating,setspace,latexsym,amsmath,epsf,amssymb,amsfonts,bm,theorem,subfigure,epstopdf}
\usepackage{cite,authblk}
\usepackage{bbm}
\usepackage{algorithm}
\usepackage[noend]{algpseudocode}
\usepackage{color}
\usepackage{mathtools}

\algrenewcommand\algorithmicforall{\textbf{foreach}}
\algrenewcommand\algorithmicindent{.8em}

\newtheorem{lemma}{Lemma}

\newenvironment{Proof}[1]{\medskip\par\noindent{\bf Proof:\,}\,#1}{{\mbox{\,$\blacksquare$}\par}}

\IEEEoverridecommandlockouts
\allowdisplaybreaks

\begin{document}

\title{Timely Tracking of Infection Status of \\ Individuals in a Population \thanks{This work was supported by NSF Grants CCF 17-13977 and ECCS 18-07348.}}

\author{Melih Bastopcu \qquad Sennur Ulukus\\
	\normalsize Department of Electrical and Computer Engineering\\
	\normalsize University of Maryland, College Park, MD 20742\\
	\normalsize \emph{bastopcu@umd.edu} \qquad \emph{ulukus@umd.edu}}

\maketitle

\begin{abstract}
We consider real-time timely tracking of infection status (e.g., covid-19) of individuals in a population. In this work, a health care provider wants to detect infected people as well as people who recovered from the disease as quickly as possible. In order to measure the timeliness of the tracking process, we use the long-term average difference between the actual infection status of the people and their real-time estimate by the health care provider based on the most recent test results. We first find an analytical expression for this average difference for given test rates, and given infection and recovery rates of people. Next, we propose an alternating minimization based algorithm to minimize this average difference. We observe that if the total test rate is limited, instead of testing all members of the population equally, only a portion of the population is tested based on their infection and recovery rates. We also observe that increasing the total test rate helps track the infection status better. In addition, an increased population size increases diversity of people with different infection and recovery rates, which may be exploited to spend testing capacity more efficiently, thereby improving the system performance. Finally, depending on the health care provider's preferences, test rate allocation can be altered to detect either the infected people or the recovered people more quickly.    
\end{abstract}

\section{Introduction}
We consider the problem of timely tracking of an infectious disease, e.g., covid-19, in a population of $n$ people. In this problem, a health care provider wants to detect infected people as quickly as possible in order to take precautions such as isolating them from the rest of the population. The health care provider also wants to detect people who recovered from the disease as soon as possible since these people need to return to work which is especially critical in sectors such as health care, food retail, and public transportation. Ideally, the health care provider should test all people all the time. However, as the total test rate is limited, the question is how frequently the health care provider should apply tests on these people when their infection and recovery rates are known. In a broader sense, this problem is related to timely tracking of multiple processes in a resource-constrained setting where each process takes binary values of $0$ and $1$ with different change rates.

\begin{figure}[t]
	\centerline{\includegraphics[width=0.9\columnwidth]{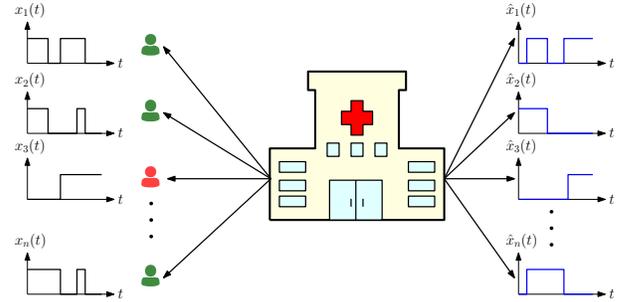}}
	\vspace{-0.1cm}
	\caption{System model. There are $n$ people whose infection status are given by $x_i(t)$. The health care provider applies tests on these people. Based on the test results, estimations for the infection status $\hat{x}_i(t)$ are generated. Infected people are shown in red color and healthy people are shown in green color.}
	\label{Fig:system_model}
	\vspace{-0.5cm}
\end{figure}

Recent studies have shown that people who recovered from infectious diseases such as covid-19 can be reinfected. Furthermore, the recovery times of individuals from the disease may vary significantly. For these reasons, in this problem, the $i$th person gets infected with rate $\lambda_i$ which is independent of the others. Similarly, the $i$th person recovers from the disease with rate $\mu_i$.\footnote{We note that the index $i$ may represent a specific individual or a group of individuals that have common features such as age, gender, profession. For example, $i=1$ may denote men between ages 70-75 who live in nursing homes, and $i=2$ may denote women between ages of 20-25 who work in the  medical field, and so on. Therefore, depending on the demographics, coefficients $\lambda_i$ and $\mu_i$ may be statistically known by the health care provider.} We denote the infection status of the $i$th person as $x_i(t)$ (shown with the black curves on the left in Fig.~\ref{Fig:system_model}) which takes the value 1 when the person is infected and the value 0 when the person is healthy. The health care provider applies tests to people marked as healthy with rate $s_i$ and to people marked as infected with rate $c_i$. Based on the test results, the health care provider forms an estimate for the infection status of the $i$th person denoted by $\hat{x}_{i}(t)$ (shown with the blue curves on the right in Fig.~\ref{Fig:system_model}) which takes the value 1 when the most recent test result is positive and the value 0, otherwise. 

We measure the timeliness of the tracking process by the difference between the actual infection status of people and the real-time estimate of the health care provider which is based on the most recent test results. We note that the difference can occur in two different cases: i) when the person is sick ($x_i(t) =1$) and the health care provider maps this person as healthy ($\hat{x}_i(t) =0$), and ii) when the person recovers from the disease ($x_i(t) =0$) but the health care provider still considers this person as infected ($\hat{x}_i(t) =1$). The former case represents the error due to late detection of infected people, while the latter case represents the error due to late detection of healed people. Depending on the health care provider's preferences, detecting infected people may be more important than detecting recovered people, or vice versa.  

Age of information has been proposed to measure timeliness of information in communication systems, and studied in the context of queueing networks, caching systems, energy harvesting systems, scheduling in networks, multi-hop multicast networks, remote estimation, lossless and lossy source coding, computation-intensive systems, vehicular, IoT, UAV systems, and so on \cite{ Najm17, Soysal18, Yates17b, Bastopcu20d, Farazi18, Wu18, Ayan19, Baknina18b, Leng19, Arafa19c, Gu20, Arafa20a, Elmagid19, bastopcu_soft_updates_journal, Bastopcu20e, Ceran18, Yates17a, Kadota18a, Hsu18b, Buyukates19b, Buyukates18b, Wang19a, Bastopcu20a, Sun17b, Yun18, Kam20, Chakravorty18, Mayekar20, MelihBatu4, Ramirez19, Buyukates19c, Zou19a, Bastopcu19, Rajaraman18, Bedewy19, Banerjee20}. Most relevant to our work, the real-time timely estimation of a single and multiple counting processes \cite{Wang19a, Bastopcu20a}, a Wiener process \cite{Sun17b}, a random walk process \cite{Yun18}, a binary Markov source \cite{Kam20} have been studied. The work that is closest to our work is reference \cite{Kam20} where the remote estimation of a symmetric binary Markov source is studied in a time-slotted system by finding the optimal sampling policies via formulating a Markov Decision Process (MDP) for real-time error, AoI and AoII metrics. Different from \cite{Kam20}, in our work, we consider real-time timely estimation of multiple non-symmetric binary sources for a continuous time system. We note that in our work, the sampler (the health care provider) does not know the states of the sources (infection status of people), and thus takes the samples (applies medical tests) randomly with fixed rates. Thus, in our work, we optimize the test rates of people to minimize the real-time estimation error.          

In this paper, we consider the real-time timely tracking of infection status of $n$ people. We first find an analytical expression for the long-term average difference between the actual infection status of people and the estimate of the health care provider based on test results. Then, we propose an alternating minimization based algorithm to find the test rates $s_i$ and $c_i$ for all people. We observe that if the total test rate is limited, we may not apply tests on all people equally. Increasing the total test rate helps track the infection status of people better, and increasing the size of the population increases diversity which may be exploited to improve the performance. Finally, depending on the health care provider's priorities, we can allocate more tests to people marked as healthy to detect the infections more quickly or to people marked as infected to detect the recoveries more quickly.        

\section{System Model} \label{sect:system_model}
\begin{figure}[t]
\begin{center}
\subfigure[]
{\includegraphics[width=0.95\linewidth]{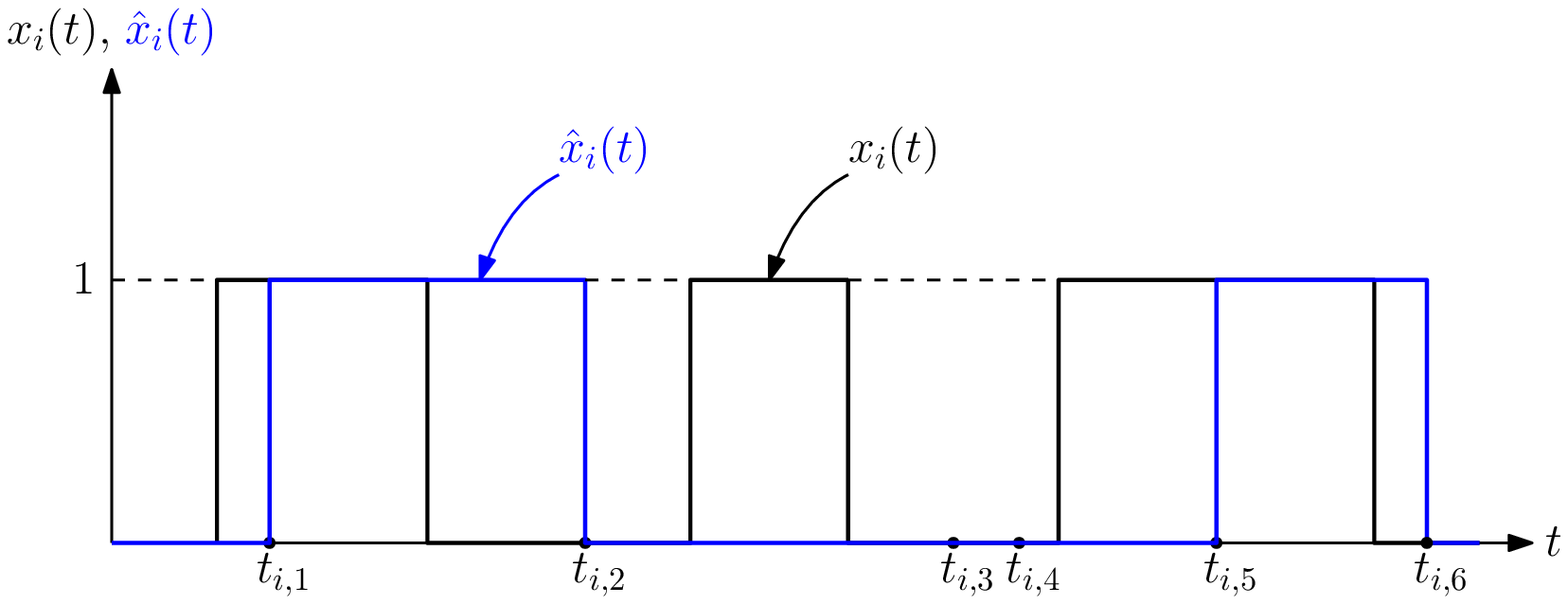}}\\ \vspace{-0.35cm}
\subfigure[]
{\includegraphics[width=0.95\linewidth]{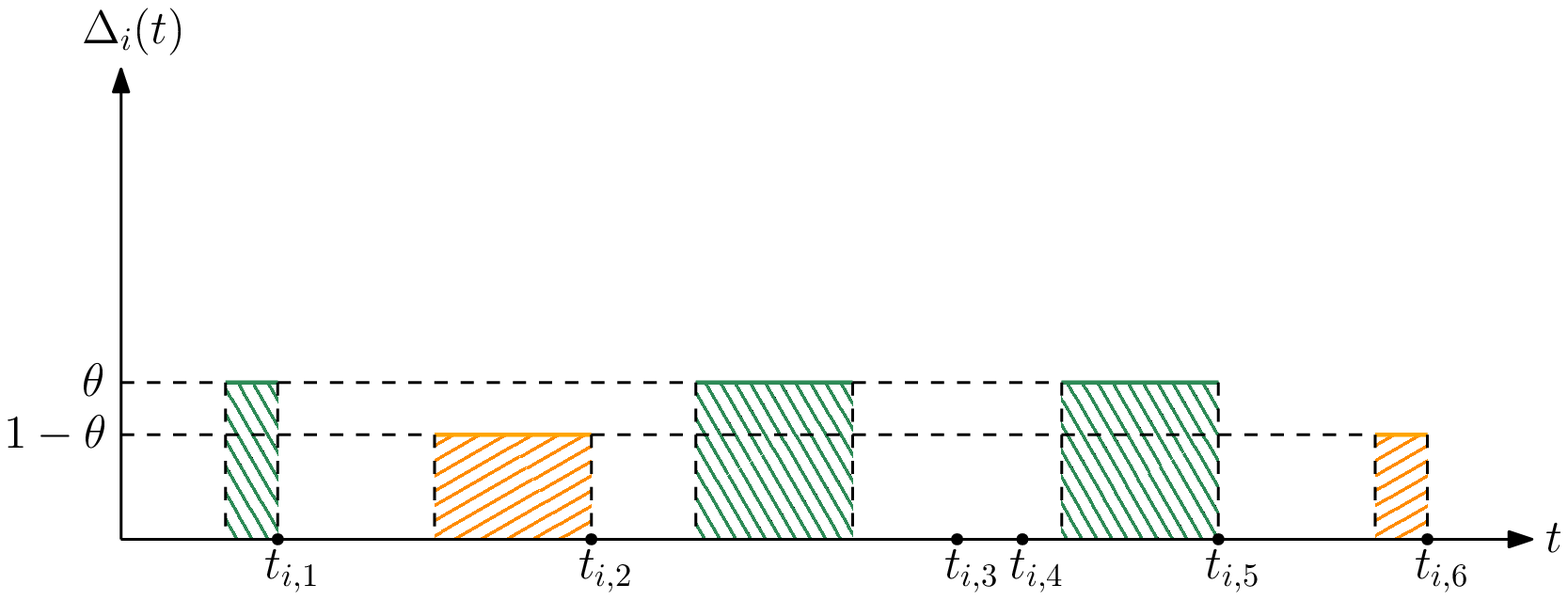}}
\end{center}
\vspace{-0.35cm}
\caption{(a) A sample evolution of $x_i(t)$ and $\hat{x}_i(t)$, and (b) the corresponding $\Delta_i(t)$ in (\ref{delta_t}). Green areas correspond to the error caused by $\Delta_{i1}(t)$ in (\ref{delta_i1}). Orange areas correspond to the error caused by $\Delta_{i2}(t)$ in (\ref{delta_i2}).}
\label{fig:model}
\vspace{-0.5cm}
\end{figure}

We consider a population of $n$ people. We denote the infection status of the $i$th person at time $t$ as $x_i(t)$ (black curve in Fig.~\ref{fig:model}(a)) which takes binary values $0$ or $1$ as follows,
\begin{align}
x_{i}(t) = \begin{cases} 
1, & \text{if the $i$th person is infected at time $t$}, \\
0, & \text{otherwise}.
\end{cases}
\end{align}

In this paper, we consider a model where each person can be infected multiple times after recovering from the disease. We denote the time interval that the $i$th person stays healthy for the $j$th time as $W_i(j)$ which is exponentially distributed with rate $\lambda_i$. We denote the recovery time for the $i$th person after infected with the virus for the $j$th time as $R_i(j)$ which is exponentially distributed with rate $\mu_i$.

A health care provider wants to track the infection status of each person. Based on the test results at times $t_{i,\ell}$, the health care provider generates an estimate for the status of the $i$th person denoted as $\hat{x}_i(t)$ (blue curve in Fig.~\ref{fig:model}(a)) by 
\begin{align}
\hat{x}_{i}(t) = x_i(t_{i,\ell}), \quad t_{i,\ell}\leq t<t_{i,\ell+1}.
\end{align}
When $\hat{x}_{i}(t)$ is 1, the health care provider applies the next test to the $i$th person after an exponentially distributed time with rate $c_i$. When $\hat{x}_{i}(t)$ is 0, the next test is applied to the $i$th person after an exponentially distributed time with rate $s_i$. 

An estimation error happens when the actual infection status of the $i$th person, $x_i(t)$, is different than the estimate of the health care provider, $\hat{x}_{i}(t)$, at time $t$. This could happen in two ways: when $x_i(t) = 1$ and $\hat{x}_{i}(t) = 0$, i.e., when the $i$th person is sick, but it has not been detected by the health care provider, and when $x_i(t) = 0$ and $\hat{x}_{i}(t) = 1$, i.e., when the $i$th person has recovered, but the health care provider does not know that the $i$th person has recovered. 

We denote the error caused by the former case, i.e., when $x_i(t) = 1$ and $\hat{x}_{i}(t) = 0$, by $\Delta_{i1}(t)$ (green areas in Fig.~\ref{fig:model}(b)), 
\begin{align}\label{delta_i1}
\Delta_{i1}(t) = \max \{x_i(t)- \hat{x}_{i}(t), 0\},
\end{align}
and we denote the error caused by the latter case, i.e., when $x_i(t) = 0$ and $\hat{x}_{i}(t) = 1$, by $\Delta_{i2}(t)$ (orange areas in Fig.~\ref{fig:model}(b)), 
\begin{align}\label{delta_i2}
\Delta_{i2}(t) = \max \{\hat{x}_{i}(t)-x_i(t), 0\}.
\end{align}
Then, the total estimation error for the $i$th person $\Delta_i(t)$ is 
\begin{align}\label{delta_t}
\Delta_{i}(t) = \theta \Delta_{i1}(t) + (1-\theta)\Delta_{i2}(t), 
\end{align}
where $\theta$ is the importance factor in $[0,1]$. A large $\theta$ gives more importance to the detection of infected people, and a small $\theta$ gives more importance to the detection of recovered people. 

We define the long-term weighted average difference between $x_i(t)$ and $\hat{x}_i(t)$ as  
\begin{align}\label{long_term}
   \Delta_{i} = \lim_{T\to\infty} \frac{1}{T}\int_0^T \Delta_{i}(t)dt.
\end{align}
Then, the overall average difference of all people $\Delta$ is
\begin{align}\label{total_age}
   \Delta = \frac{1}{n} \sum_{i=1}^{n} \Delta_{i}.
\end{align}

Our aim is to track the infection status of all people. Due to limited resources, there is a total test rate constraint $\sum_{i=1}^{n}s_i+\sum_{i=1}^{n}c_i\leq C$. Thus, our aim is to find the optimal test rates $s_i$ and $c_i$ to minimize $\Delta$ in (\ref{total_age}) while satisfying this total test rate constraint. We formulate the following problem,
\begin{align}
\label{problem1}
\min_{\{s_i, c_i \}}  \quad &  \Delta \nonumber \\
\mbox{s.t.} \quad & \sum_{i=1}^{n} s_i + \sum_{i=1}^{n} c_i\leq C \nonumber \\
\quad & s_i\geq 0, \quad c_i\geq 0,\quad i=1,\dots,n.
\end{align} 
In the next section, we find the total average difference $\Delta$. 

\begin{figure}[t]
\begin{center}
\subfigure[]
{\includegraphics[width=0.95\linewidth]{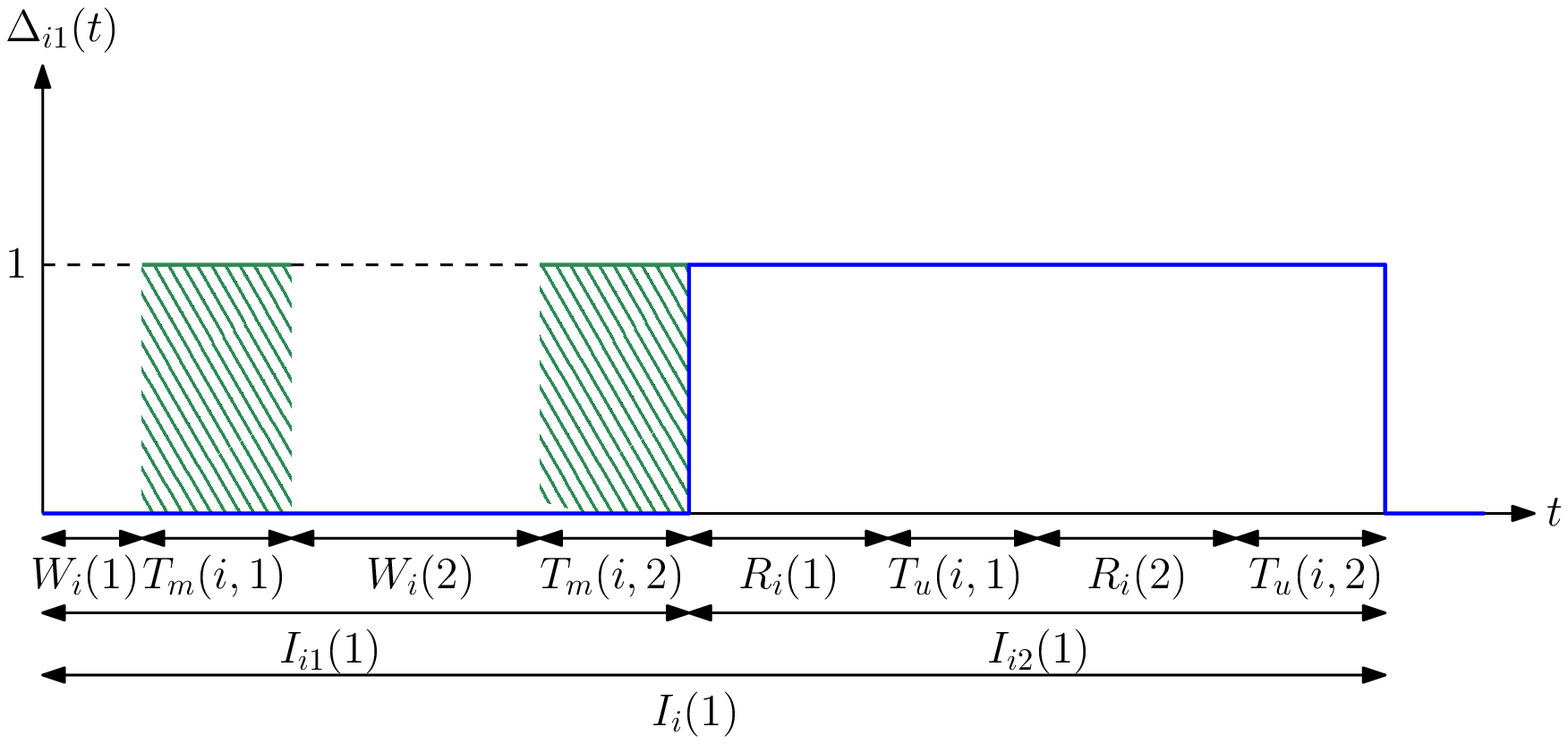}}\\ \vspace{-0.35cm}
\subfigure[]
{\includegraphics[width=0.95\linewidth]{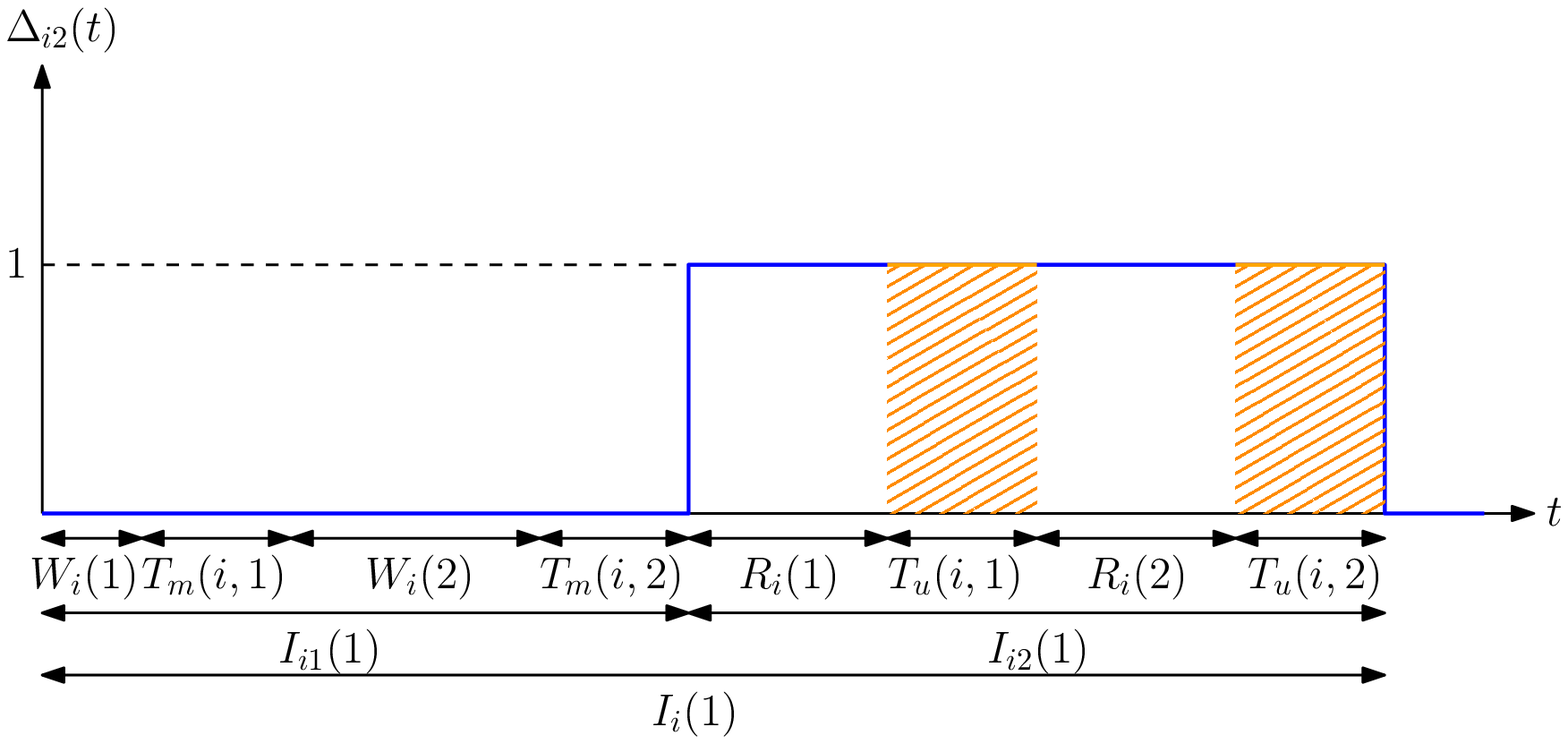}}
\end{center}
\vspace{-0.35cm}
\caption{A sample evolution of (a) $\Delta_{i1}(t)$, and (b) $\Delta_{i2}(t)$ in a typical cycle.}
\label{fig:error}
\vspace{-0.5cm}
\end{figure}

\section{Average Difference Analysis} \label{Sec:Average_difference} 
We first find analytical expressions for $\Delta_{i1}(t)$ in (\ref{delta_i1}) and $\Delta_{i2}(t)$ in (\ref{delta_i2}). 
We note that $\Delta_{i1}(t)$ can be equal to $1$ when $\hat{x}_i(t) = 0$ and is always equal to $0$ when $\hat{x}_i(t) = 1$. Assume that at time $0$, both $x_i(0)$ and $\hat{x}_i(0)$ are 0. After an exponentially distributed time with rate $\lambda_i$, which is denoted by $W_i$, the $i$th person is infected, and thus $x_i(t)$ becomes $1$. At that time, since $\hat{x}_i(t) = 0$, $\Delta_{i1}(t)$ becomes $1$. $\Delta_{i1}(t)$ will be equal to 0 again either when the $i$th person recovers from the disease which happens after $R_i$ which is exponentially distributed with rate $\mu_i$ or when the health care provider performs a test on the $i$th person after $D_i$ which is exponentially distributed with rate $s_i$. We define $T_m(i)$ as the earliest time at which one of these two cases happens, i.e., $T_m(i) = \min \{R_i,D_i \}$. We note that $T_m(i)$ is also exponentially distributed with rate $ \mu_i+s_i$, and we have $\mathbb{P} (T_m(i) = R_i) = \frac{\mu_i}{\mu_i+s_i}$ and $\mathbb{P} (T_m(i) = D_i) = \frac{s_i}{\mu_i+s_i}$. If the $i$th person recovers from the disease before testing, we return to the initial case where both $x_i(t)$ and $\hat{x}_i(t)$ are equal to $0$ again. In this case, this cycle repeats itself, i.e., the $i$th person becomes sick again after $W_i$ and $\Delta_{i1}(t)$ remains as 1 until either the person recovers or the health care provider performs a test which takes another $T_m(i)$ duration. If the health care provider performs a test before the person recovers, then $\hat{x}_i(t)$ becomes $1$. We denote the time interval for which $\hat{x}_i(t)$ stays at 0 as $I_{i1}$ which is given by 
\begin{align}
    I_{i1} = \sum_{\ell = 1}^{K_1} T_m(i,\ell)+ W_i(\ell), 
\end{align}
where $K_1$ is geometric with rate $\mathbb{P} (T_m(i) = D_i) = \frac{s_i}{\mu_i+s_i}$. Due to \cite[Prob. 9.4.1]{Yates14}, $\sum_{\ell = 1}^{K_1} T_m(i,\ell)$ and $\sum_{\ell = 1}^{K_1} W_i(\ell)$ are exponentially distributed with rates $s_i$ and $\frac{\lambda_i s_i}{\mu_i+s_i}$, respectively. As $\mathbb{E}[I_{i1}] = \mathbb{E}[\sum_{\ell = 1}^{K_1} T_m(i,\ell)]+\mathbb{E}[\sum_{\ell = 1}^{K_1} W_i(\ell)]$, we have
\begin{align}\label{I_1}
\mathbb{E}[I_{i1}] = \frac{1}{s_i}+\frac{s_i+\mu_i}{s_i\lambda_i }.
\end{align}

When $\hat{x}_i(t) =1$, the health care provider marks the $i$th person as infected. The $i$th person recovers from the virus after $R_i$. After the $i$th person recovers, either the health care provider performs a test after $Z_i$  which is exponentially distributed with rate $c_i$ or the $i$th person is reinfected with the virus which takes $W_i$ time. We define $T_u(i)$ as the earliest time at which one of these two cases happens, i.e., $T_u(i) = \min \{W_i, Z_i\}$. Similarly, we note that $T_u(i)$ is exponentially distributed with rate $\lambda_i+c_i$, and we have $\mathbb{P} (T_u(i) = W_i) = \frac{\lambda_i}{\lambda_i+c_i}$ and $\mathbb{P} (T_m(i) = Z_i) = \frac{c_i}{\lambda_i+c_i}$. If the person is reinfected with the virus before a test is applied, this cycle repeats itself, i.e., the $i$th person recovers after another $R_i$, and then either a test is applied to the $i$th person, or the person is infected again which takes another $T_u(i)$. If the health care provider performs a test to the $i$th person before the person is reinfected, the health care provider marks the $i$th person as healthy again, i.e., $\hat{x}_i(t)$ becomes 0. We denote the time interval that $\hat{x}_i(t)$ is equal to 1 as $I_{i2}$ which is given by
\begin{align}
    I_{i2} = \sum_{\ell = 1}^{K_2} T_u(i,\ell)+ R_i(\ell), 
\end{align}
where $K_2$ is geometric with rate $\mathbb{P} (T_u(i) = Z_i) = \frac{c_i}{\lambda_i+c_i}$. Similarly, $\sum_{\ell = 1}^{K_2} T_u(i,\ell)$ and $\sum_{\ell = 1}^{K_2} R_i(\ell)$ are exponentially distributed with rates $c_i$ and $\frac{c_i\mu_i}{\lambda_i+c_i}$, respectively. As $\mathbb{E}[I_{i2}] = \mathbb{E}[\sum_{\ell = 1}^{K_2} T_u(i,\ell)]+\mathbb{E}[\sum_{\ell = 1}^{K_2} R_i(\ell)]$, we have
\begin{align}\label{I_2}
    \mathbb{E}[I_{i2}] = \frac{1}{c_i}+\frac{c_i+\lambda_i}{c_i\mu_i }.
\end{align}

We denote the time interval between the $j$th and $(j+1)$th times that $\hat{x}_i(t)$ changes from 1 to 0 as the $j$th cycle $I_i(j)$ where $I_i(j) =I_{i1}(j)+ I_{i2}(j)$. We note that $\Delta_{i1}(t)$ is always equal to 0 during $I_{i2}(j)$, i.e., $\hat{x}_i(t)= 1$, and $\Delta_{i1}(t)$ is equal to 1 when $x_i(t) = 1$ in $I_{i1}(j)$. We denote the total time duration when $\Delta_{i1}(t)$ is equal to 1 as $T_{e,1}(i,j)$ during the $j$th cycle where $T_{e,1}(i,j) =\sum_{\ell = 1}^{K_1} T_m(i,\ell)$. Thus, we have $\mathbb{E}[T_{e,1}(i)] = \frac{1}{s_i}$. Then, using ergodicity, similar to \cite{Bastopcu20d}, $\Delta_{i1}$ is equal to
\begin{align}\label{Delta_i1}
\Delta_{i1}  = \frac{\mathbb{E}[T_{e,1}(i)]}{\mathbb{E}[I_{i}]}= \frac{\mathbb{E}[T_{e,1}(i)]}{\mathbb{E}[I_{i1}]+\mathbb{E}[I_{i2}]}.
\end{align}
Thus, we have 
\begin{align}\label{Delta_i1_val}
\Delta_{i1} = \frac{\mu_i \lambda_i}{\mu_i+\lambda_i}\frac{c_i }{\mu_ic_i+\lambda_i s_i+c_is_i}.
\end{align}

Next, we find $\Delta_{i2}$. We note that $\Delta_{i2}(t)$ is equal to 1 when $x_i(t) = 0$ in $I_{i2}(j)$ and is always equal to 0 during $I_{i1}(j)$. Similarly, we denote the total time duration where $\Delta_{i2}(t)$ is equal to 1 in the $j$th cycle $I_i(j)$ as $T_{e,2}(i,j)$ which is equal to $T_{e,2}(i,j) =\sum_{\ell = 1}^{K_2} T_u(i,\ell)$. Thus, we have $\mathbb{E}[T_{e,2}(i)] = \frac{1}{c_i}$. Then, similar to $\Delta_{i1}$ in (\ref{Delta_i1}), $\Delta_{i2}$ is equal to     
\begin{align}\label{Delta_i2_val}
\Delta_{i2} =\frac{\mu_i \lambda_i}{\mu_i+\lambda_i}\frac{s_i}{\mu_ic_i+\lambda_i s_i+c_is_i}.
\end{align}

By using (\ref{delta_t}), (\ref{Delta_i1_val}), and (\ref{Delta_i2_val}), we obtain $\Delta_i$ as
\begin{align}\label{delta_i}
    \Delta_i = \frac{\mu_i \lambda_i}{\mu_i+\lambda_i}\frac{\theta c_i + (1-\theta)s_i}{\mu_ic_i+\lambda_i s_i+c_is_i}.
\end{align}
Then, by inserting (\ref{delta_i}) in (\ref{total_age}), we obtain $\Delta$. In the next section, we solve the optimization problem in (\ref{problem1}). 

\section{Optimization of Average Difference} \label{sect:opt_soln}
In this section, we solve the optimization problem in (\ref{problem1}). 
Using $\Delta_i$ in (\ref{delta_i}) in (\ref{total_age}), we rewrite (\ref{problem1}) as
\begin{align}
\label{problem1_opt}
\min_{\{s_i, c_i \}}  \quad &  \sum_{i=1}^{n} \frac{\mu_i \lambda_i}{\mu_i+\lambda_i}\frac{\theta c_i + (1-\theta)s_i}{\mu_i c_i+\lambda_i s_i+c_is_i} \nonumber \\
\mbox{s.t.} \quad & \sum_{i=1}^{n} s_i + \sum_{i=1}^{n} c_i\leq C \nonumber \\
\quad & s_i\geq 0, \quad c_i\geq 0,\quad i=1,\dots,n,
\end{align}
We define the Lagrangian function \cite{Boyd04} for (\ref{problem1_opt}) as
\begin{align}\label{lagrange}
    \mathcal{L} =& \sum_{i=1}^{n} \frac{\mu_i \lambda_i}{\mu_i+\lambda_i}\frac{\theta c_i + (1-\theta)s_i}{\mu_i c_i+\lambda_i s_i+c_is_i}+\beta \left(\sum_{i=1}^{n} s_i + c_i- C \right)\nonumber\\
    &-\sum_{i=1}^{n}\nu_i s_i -\sum_{i=1}^{n}\eta_i c_i, 
\end{align}
where $\beta \geq 0$, $\nu_i\geq 0$, and $ \eta_i\geq 0$. The KKT conditions are
\begin{align}
\frac{\partial \mathcal{L}}{\partial s_i} =& \frac{\mu_i \lambda_i c_i}{\mu_i+\lambda_i}\frac{(1-\theta)\mu_i -\theta(c_i + \lambda_i)}{(\mu_i c_i+\lambda_i s_i + s_i c_i)^2}+\beta-\nu_i = 0,\label{KKT1}\\
    \frac{\partial \mathcal{L}}{\partial c_i} =& \frac{\mu_i \lambda_i s_i}{\mu_i+\lambda_i}\frac{\theta \lambda_i -(1-\theta)(\mu_i + s_i)}{(\mu_i c_i+\lambda_i s_i + s_i c_i)^2}+\beta-\eta_i = 0,\label{KKT2}
\end{align}
for all $i$. The complementary slackness conditions are 
\begin{align}
    \beta \left(\sum_{i=1}^{n} s_i + c_i- C\right)  = 0,\quad \nu_i s_i = 0, \quad \eta_i c_i =0.\label{CS}
\end{align}

First, we find $s_i$. From (\ref{KKT1}), we have
\begin{align}\label{temp_si}
    (\mu_i c_i+\lambda_i s_i + s_i c_i)^2= \frac{\mu_i \lambda_i c_i}{\mu_i+\lambda_i}\frac{\theta(c_i + \lambda_i)-(1-\theta)\mu_i }{\beta-\nu_i}.
\end{align}
When $\theta(c_i + \lambda_i)\geq (1-\theta)\mu_i$, we solve (\ref{temp_si}) for $s_i$ as 
\begin{align}\label{soln_si}
    s_i = \frac{\mu_i c_i}{\lambda_i+c_i} \left(\!\sqrt{\frac{1}{\mu_i c_i}\frac{\lambda_i}{\mu_i +\lambda_i}\frac{\theta(c_i + \lambda_i)-(1-\theta)\mu_i}{\beta}} -1\!\right)^+\!,
\end{align}
where we used the fact that we either have $s_i>0$ and $\nu_i = 0$, or $s_i = 0$ and $\nu_i\geq 0$, due to (\ref{CS}). Here, $(\cdot)^+ = \max(\cdot,0)$.

Finally, when $\theta(c_i + \lambda_i)< (1-\theta)\mu_i$, we have $\frac{\partial \Delta_i}{\partial s_i}> 0$, and thus it is optimal to choose $s_i = 0$ as our aim is to minimize $\Delta$ in (\ref{total_age}). In this case, when $s_i = 0$, we have $\Delta_i = \frac{\theta \lambda_i}{\mu_i+\lambda_i}$ which is independent of the value of $c_i$. As we obtain the same $\Delta_i$ for all values of $c_i$, and the total update rate is limited, i.e., $\sum_{i=1}^{n} s_i + c_i \leq  C$, in this case, it is optimal to choose $c_i = 0$ as well (i.e., when $s_i =0$). 

Next, we find $c_i$. From (\ref{KKT2}), we have
\begin{align}\label{ci_eqn}
 (\mu_i c_i+\lambda_i s_i + s_i c_i)^2 = \frac{\mu_i \lambda_i s_i}{\mu_i+\lambda_i}\frac{(1-\theta)(\mu_i + s_i)-\theta \lambda_i }{\beta-\eta_i}.
\end{align}
When $(1-\theta)(\mu_i + s_i)\geq \theta \lambda_i$, we solve (\ref{ci_eqn}) for $c_i$ as 
\begin{align}\label{soln_ci}
c_i = \frac{\lambda_i s_i}{\mu_i + s_i}\left(\!\sqrt{\frac{1}{\lambda_i s_i}\frac{\mu_i}{\mu_i+\lambda_i}\frac{(1-\theta)(s_i+\mu_i)-\theta \lambda_i}{\beta}}-1 \!\right)^+\!,
\end{align}
where we used the fact that we either have $c_i>0$ and $\eta_i = 0$, or $c_i = 0$ and $\eta_i\geq 0$, due to (\ref{CS}).

Similarly, when $(1-\theta)(s_i+\mu_i)< \theta \lambda_i$, we have $\frac{\partial \Delta_i}{\partial c_i} >0$. Thus, in this case, it is optimal to choose $c_i = 0$. When $c_i = 0$, we have $\Delta_i = \frac{(1-\theta)\mu_i}{\mu_i+\lambda_i}$ which is independent of the value of $s_i$. Thus, it is optimal to choose $s_i = 0$ when $c_i = 0$.       

From (\ref{soln_si}), if $\frac{1}{\mu_i c_i}\frac{\lambda_i}{\mu_i+\lambda_i}(\theta(c_i + \lambda_i)-(1-\theta)\mu_i)\leq \beta$, we must have $s_i = 0$. Thus, for a given $c_i$, the optimal test rate allocation policy for $s_i$ is a threshold policy where $s_i$'s with small $\frac{1}{\mu_i c_i}\frac{\lambda_i}{\mu_i+\lambda_i}(\theta(c_i + \lambda_i)-(1-\theta)\mu_i)$ are equal to zero. Similarly, from (\ref{soln_ci}), if $ \frac{1}{\lambda_i s_i}\frac{\mu_i}{\mu_i+\lambda_i}\left((1-\theta)(s_i+\mu_i)-\theta \lambda_i\right)\leq \beta$, we must have $c_i = 0$. Thus, for a given $s_i$, the optimal policy to determine $c_i$ is a threshold policy where $c_i$'s with small $ \frac{1}{\lambda_i s_i}\frac{\mu_i}{\mu_i+\lambda_i}\left((1-\theta)(s_i+\mu_i)-\theta \lambda_i\right)$ are equal to zero. 

Next, we show that in the optimal policy, if $s_i>0$ and $c_i>0$ for some $i$, then the total test rate constraint must be satisfied with equality, i.e., $\sum_{i=1}^{n} s_i+c_i = C$.

\begin{lemma}\label{lemma1}
    In the optimal policy, if $s_i>0$ and $c_i>0$ for some $i$, then we have $\sum_{i=1}^{n} s_i+c_i = C$.
\end{lemma}

\begin{Proof}
The derivatives of $\Delta_i$ with respect to $s_i$ and $c_i$ are  
\begin{align}
    \frac{\partial \Delta_i}{\partial s_i} = \frac{\mu_i \lambda_i c_i}{\mu_i + \lambda_i}\frac{(1-\theta)\mu_i-\theta(c_i+\lambda_i)}{\left(c_i \mu_i +s_i c_i+ \lambda_i s_i\right)^2},\\
    \frac{\partial \Delta_i}{\partial c_i} = \frac{\mu_i \lambda_i s_i}{\mu_i + \lambda_i}\frac{\theta\lambda_i-(1-\theta)(s_i+\mu_i)}{\left(c_i \mu_i +s_i c_i+ \lambda_i s_i\right)^2}.
\end{align}
We note that $s_i>0$ in (\ref{soln_si}) implies that $\theta(c_i+\lambda_i) > (1-\theta)\mu_i$. In this case, we have $\frac{\partial \Delta_i}{\partial s_i} <0$. Similarly, $c_i>0$ in (\ref{soln_ci}) implies that $(1-\theta)(s_i+\mu_i)>\theta \lambda_i$. Thus, we have $\frac{\partial \Delta_i}{\partial c_i}<0$. Therefore, in the optimal policy, if we have $s_i>0$ and $c_i>0$ for some $i$, then we must have $\sum_{i=1}^{n} s_i+c_i = C$. Otherwise, we can further decrease $\Delta$ in (\ref{total_age}) by increasing $c_i$ or $s_i$.        
\end{Proof}

Next, we propose an alternating minimization based algorithm for finding $s_i$ and $c_i$. For this purpose, for given initial $(s_i, c_i)$ pairs, we define $\phi_i$ as 
\begin{align} \label{phi-i-defn}
\phi_i \!= \! \begin{cases} 
\frac{1}{\mu_i c_i}\frac{\lambda_i}{\mu_i +\lambda_i}(\theta(c_i + \lambda_i)-(1-\theta)\mu_i), \hspace{1mm} i\!=\!1,\dots,n, \\
\frac{1}{\lambda_i s_i}\frac{\mu_i}{\mu_i+\lambda_i}((1-\theta)(s_i+\mu_i)-\theta \lambda_i), \hspace{1mm} i \!= \!n+1, \dots, 2n.
\end{cases}
\end{align}
Then, we define $u_i$ as
\begin{align}\label{eqn_ui}
 u_i =  \begin{cases} 
\frac{\mu_i c_i}{\lambda_i+c_i}\left( \sqrt{\frac{\phi_i}{\beta}}-1\right)^+, & i=1,\dots,n, \\
\frac{\lambda_i s_i}{\mu_i +s_i}\left(\sqrt{\frac{\phi_i}{\beta}}-1 \right)^+, & i = n+1, \dots, 2n.
\end{cases}   
\end{align}
From (\ref{soln_si}) and (\ref{soln_ci}), $s_i = u_i$ and $c_i = u_{n+i}$, for $i = 1,\dots,n$.

Next, we find $s_i$ and $c_i$ by determining $\beta$ in (\ref{eqn_ui}). First, assume that, in the optimal policy, there is an $i$ such that $s_i>0$ and $c_i>0$. Thus, by Lemma~\ref{lemma1}, we must have $\sum_{i=1}^{n}s_i+c_i = C$. We initially take random $(s_i, c_i)$ pairs such that $\sum_{i=1}^{n}s_i+c_i = C$. Then, given the initial $(s_i, c_i)$ pairs, we immediately choose $u_i =0$ for $\phi_i<0$. For the remaining $u_i$ with $\phi_i\geq0$, we apply a solution method similar to that in \cite{Bastopcu20d}. By assuming $\phi_i\geq \beta$, i.e., by disregarding $(\cdot)^+$ in (\ref{eqn_ui}), we solve $\sum_{i=1}^{2n}u_i = C$ for $\beta$. Then, we compare the smallest $\phi_i$ which is larger than zero in (\ref{phi-i-defn}) with $\beta$. If we have $\phi_i \geq \beta$, then it implies that $u_i\geq 0 $ for all remaining $i$. Thus, we have obtained $u_i$ values for given initial ($s_i, c_i$) pairs. If the smallest $\phi_i$ which is larger than zero is smaller than $\beta$, then the corresponding $u_i$ is negative and we should choose $u_i =0$ for the smallest non-negative $\phi_i$. Then, we repeat this procedure until the smallest non-negative $\phi_i$ is larger than $\beta$. After determining all $u_i$, we obtain $s_i = u_i$ and $c_i = u_{n+i} $ for $i = 1,\dots, n$. Then, with the updated values of $(s_i, c_i)$ pairs, we keep finding $u_i$'s until the KKT conditions in (\ref{KKT1}) and (\ref{KKT2}) are satisfied. 

We note that for indices (persons) $i$ for which $(s_i,c_i)$ are zero, the health care provider does not perform any tests, and maps these people as either always infected, i.e., $\hat{x}_i(t) =1$ for all $t$, or always healthy, i.e., $\hat{x}_i(t) =0$. If $\hat{x}_i(t) =0$ for all $t$, $\Delta_i = \frac{\theta \lambda_i}{\mu_i+\lambda_i}$, and if $\hat{x}_i(t) =1$ for all $t$,  $\Delta_i = \frac{(1-\theta) \mu_i}{\mu_i+\lambda_i}$. Thus, for such $i$, the health care provider should choose $\hat{x}_i(t) =0$ for all $t$, if $\frac{\theta \lambda_i}{\mu_i+\lambda_i}< \frac{(1-\theta) \mu_i}{\mu_i+\lambda_i}$, and should choose $\hat{x}_i(t) =1$ for all $t$, otherwise, without performing any tests. 

Finally, we note that the problem in (\ref{problem1_opt}) is not a convex optimization problem as the objective function is not jointly convex in $s_i$ and $c_i$. Therefore, the solutions obtained via the proposed method may not be globally optimal. For that reason, we choose different initial starting points and apply the proposed alternating minimization based algorithm and choose the solution that achieves the smallest $\Delta$ in (\ref{total_age}).    

\section{Numerical Results} \label{sect:num_res}
In this section, we provide four numerical results. For these examples, we take $\lambda_i$ as
\begin{align}\label{lambda_i}
    \lambda_i = a r^i, \quad i=1,\dots,n,
\end{align}
where $r= 0.9$ and $a$ is such that $\sum_{i=1}^{n}\lambda_i = 6$. Also, we take $\mu_i$ as
\begin{align}\label{alpha_i}
    \mu_i = b q^i, \quad i=1,\dots,n,
\end{align}
where $q = 1.1$ and $b$ is such that $\sum_{i=1}^{n}\mu_i = 4$. Since $\lambda_i$ in (\ref{lambda_i}) decreases with $i$, people with lower indices get infected more quickly compared to people with higher indices. Since $\mu_i$ in (\ref{alpha_i}) increases with $i$, people with higher indices recover more quickly compared to people with lower indices. Thus, low index people get infected quickly and get well slowly.          

\begin{figure}[t]
	\begin{center}
	\subfigure[]{%
	\includegraphics[width=0.88\linewidth]{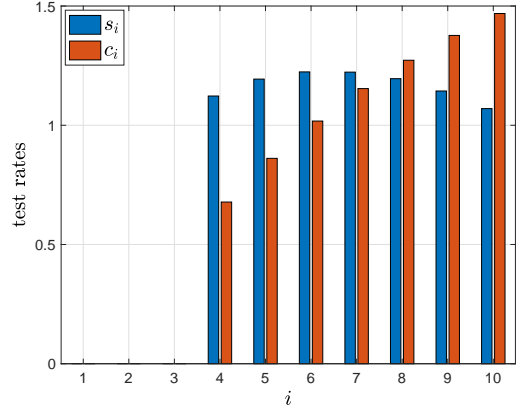}}\\\vspace{-0.35cm}
	\subfigure[]{%
	\includegraphics[width=0.88\linewidth]{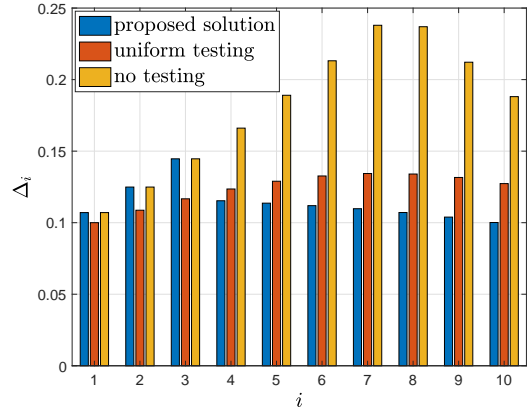}}
	\end{center}
	\vspace{-0.35cm}
	\caption{(a) Test rates $s_i$ and $c_i$, (b) corresponding average difference $\Delta_i$. }
	\label{Fig:sim1}
	\vspace{-0.5cm}
\end{figure}

In the first example, we take the total number of people as $n=10$, the total test rate as $C=16$, and $\theta = 0.5$. We start with randomly chosen $s_i$ and $c_i$ such that $\sum_{i=1}^{n}s_i+c_i =16$, and apply the alternating minimization based method proposed in Section~\ref{sect:opt_soln}. We repeat this process for 30 different initial $(s_i,c_i)$ pairs and choose the solution that gives the smallest $\Delta$. In Fig.~\ref{Fig:sim1}(a), we observe that the first three people are never tested by the health care provider. We note that $s_i$, which is the test rate when $\hat{x}_{i}(t)=0$,  initially increases with $i$ but then decreases with $i$. This means that people who get infected rarely are tested less frequently when they are marked as healthy. Similarly, we observe in Fig.~\ref{Fig:sim1}(a) that $c_i$, which is the test rate when $\hat{x}_{i}(t)=1$, monotonically increases with $i$. In other words, people who recover from the virus quickly are tested more frequently when they are marked infected.

In Fig.~\ref{Fig:sim1}(b), we plot $\Delta_i$ resulting from the solution found from the proposed algorithm, $\Delta_i$ when the health care provider applies tests to everyone in the population uniformly, i.e., $s_i = c_i=\frac{C}{2n}$  for all $i$, and $\Delta_i$ when the health care provider applies no tests, i.e., $s_i = c_i= 0$ for all $i$. In the case of no tests, we have $\Delta_i = \min\{ \frac{\theta \lambda_i}{\mu_i+\lambda_i}, \frac{(1-\theta) \mu_i}{\mu_i+\lambda_i}\}$. We observe in Fig.~\ref{Fig:sim1}(b) that the health care provider applies tests on people whose $\Delta_i$ can be reduced the most as opposed to uniform testing where everyone is tested equally. Thus, the first three people who have the smallest $\Delta_i$ are not tested by the health care provider. With the proposed solution, by not testing the first three people, $\Delta_i$ are further reduced for the remaining people compared to uniform testing. For the people who are not tested, the health care provider chooses $\hat{x}_i(t) = 1$ all the time, i.e., marks these people always sick as $\frac{\theta \lambda_i}{\mu_i+\lambda_i}> \frac{(1-\theta) \mu_i}{\mu_i+\lambda_i}$. This is expected as these people have high $\lambda_i$ and low $\mu_i$, i.e., they are infected easily and they stay sick for a long time. 

\begin{figure}[t]
	\centerline{\includegraphics[width=0.88\columnwidth]{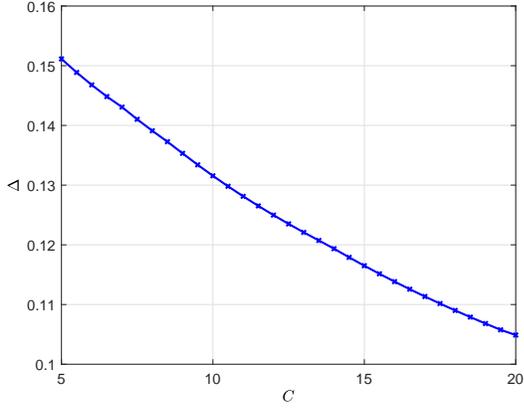}}
	\vspace{-0.35cm}
	\caption{The average difference $\Delta$ with respect to total test rate $C$.}
	\label{Fig:sim2}
	\vspace{-0.4cm}
\end{figure}

\begin{figure}[t]
	\centerline{\includegraphics[width=0.88\columnwidth]{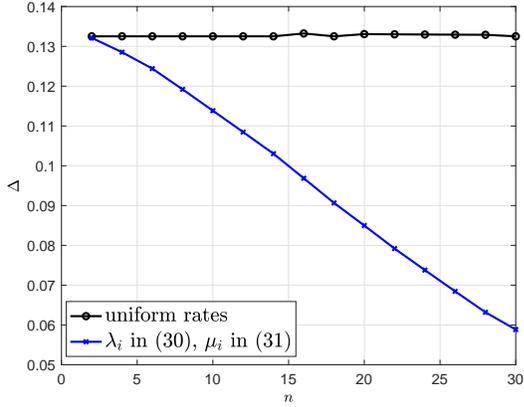}}
	\vspace{-0.3cm}
	\caption{The average difference $\Delta$ with respect to number of people $n$. We use uniform infection and healing rates, i.e., $\lambda_i = \frac{6}{n}$ and $\mu_i = \frac{4}{n}$ for all $i$, and also $\lambda_i$ in (\ref{lambda_i}) and $\mu_i$ in (\ref{alpha_i}) with $\sum_{i=1}^{n}\lambda_i = 6$ and $\sum_{i=1}^{n}\mu_i = 4$.}
	\label{Fig:sim4}
	\vspace{-0.5cm}
\end{figure}

In the second example, we use the same set of variables except for the total test rate $C$. We vary the total test rate $C$ in between $5$ and $20$. We plot $\Delta$ with respect to $C$ in Fig.~\ref{Fig:sim2}. We observe that $\Delta$ decreases with $C$. Thus, with higher total test rates, the health care provider can tract the infection status of the population better as expected.   

\begin{figure}[t]
	\begin{center}
	\subfigure[]{%
	\includegraphics[width=0.88\linewidth]{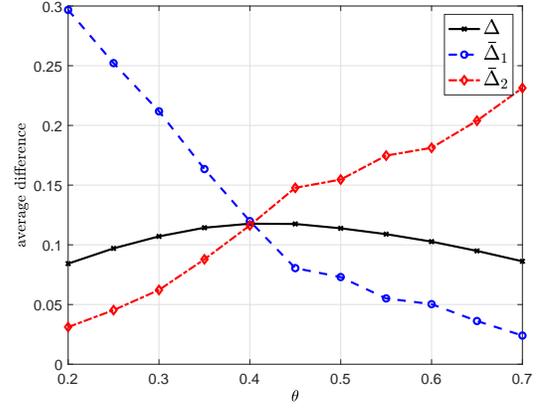}}\\\vspace{-0.35cm}
	\subfigure[]{%
	\includegraphics[width=0.88\linewidth]{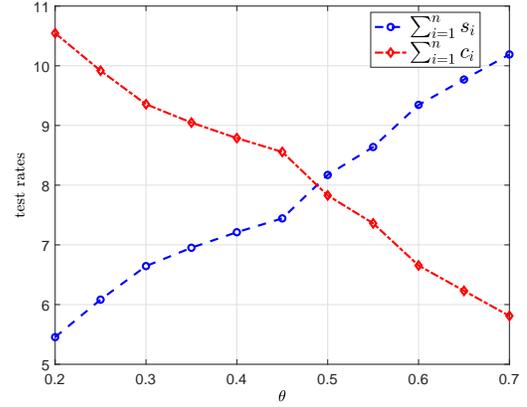}}
	\end{center}
	\vspace{-0.35cm}
	\caption{(a) $\Delta$ in (\ref{total_age}), $\bar{\Delta}_1$ which is  $\frac{1}{n}\sum_{i=1}^{n}\Delta_{i1}$, and $\bar{\Delta}_2$ which is  $\frac{1}{n}\sum_{i=1}^{n}\Delta_{i2}$, (b) corresponding total test rates $\sum_{i=1}^{n}s_{i}$ and $\sum_{i=1}^{n}c_{i}$. }
	\label{Fig:sim3}
	\vspace{-0.5cm}
\end{figure}

In the third example, we use the same set of variables except for the total number of people $n$. In addition, we also use uniform infection and healing rates, i.e., $\lambda_i = \frac{6}{n}$ and $\mu_i = \frac{4}{n}$ for all $i$, for comparison with $\lambda_i$ in (\ref{lambda_i}) and $\mu_i$ in (\ref{alpha_i}), while keeping the total infection and healing rates the same, i.e., $\sum_{i=1}^{n} \lambda_i= 6$ and $\sum_{i=1}^{n} \mu_i= 4$, for both cases. We vary the number of people $n$ from $2$ to $30$. We observe in Fig.~\ref{Fig:sim4} that when the infection and healing rates are uniform in the population, the health care provider can track the infection status with the same efficiency, even though the population size increases (while keeping the total infection and healing rates fixed). For the case of $\lambda_i$ in (\ref{lambda_i}) and $\mu_i$ in (\ref{alpha_i}), when we increase the population size, we increase the number of people who rarely get sick, i.e., people with high $i$ indices, and also people who rarely heal from the disease, i.e., people with small $i$ indices. Thus, it gets easier for the health care provider to track the infection status of the people. That is why when we use $\lambda_i$ in (\ref{lambda_i}) and $\mu_i$ in (\ref{alpha_i}), we observe in Fig.~\ref{Fig:sim4} that the health care provider can track the infection status of the people better, even though the population size increases.        

In the fourth example, we use the same set of variables as the first example except for the importance factor $\theta$. Here, we vary $\theta$ in between $0.2$ to $0.7$. We plot $\Delta$ in (\ref{total_age}), $\bar{\Delta}_{1}$ which is $\bar{\Delta}_1 = \frac{1}{n}\sum_{i=1}^{n}\Delta_{i1}$, and $\bar{\Delta}_{2}$ which is $\bar{\Delta}_2 = \frac{1}{n}\sum_{i=1}^{n}\Delta_{i2}$ in Fig.~\ref{Fig:sim3}(a). Note that $\bar{\Delta}_{1}$ represents the average difference when people are infected, but they have not been detected by the health care provider, and $\bar{\Delta}_2$ represents the average difference when people have recovered, but the health care provider still marks them as infected. Note that when $\theta$ is high, we give importance to minimization of $\bar{\Delta}_1$, i.e., the early detection of people with infection, and when $\theta$ is low, we give importance to minimization of $\bar{\Delta}_2$, i.e., the early detection of people who recovered from the disease. That is why we observe in Fig.~\ref{Fig:sim3}(a) that $\bar{\Delta}_1$ decreases with $\theta$ while $\bar{\Delta}_2$ increases with $\theta$. 

We plot the total test rates $\sum_{i=1}^{n}s_i$ and $\sum_{i=1}^{n}c_i$ in Fig.~\ref{Fig:sim3}(b). We observe in Fig.~\ref{Fig:sim3}(b) that if it is more important to detect the infected people, i.e., if $\theta$ is high, then the health care provider should apply higher test rates to people who are marked as healthy. In other words, $\sum_{i=1}^{n}s_i$ increases with $\theta$. Similarly, if it is more important to detect people who recovered from the disease, then the health care provider should apply high test rates to people who are marked as infected. That is, $\sum_{i=1}^{n}c_i$ is high when $\theta$ is low. Therefore, depending on the priorities of the health care provider, a suitable $\theta$ needs to be chosen.      
\section{Conclusion}
We considered timely tracking of infection status of individuals in a population. For exponential infection and healing processes with given rates, we determined the rates of exponential testing processes. We observed in numerical results that the test rates depend on individuals' infection and healing rates, the individuals' last known state of healthy or infected, as well as the health care provider's priorities of detecting infected people or recovered people more quickly.

\bibliographystyle{unsrt}
\bibliography{IEEEabrv,lib_v1_melih}

\begin{thebibliography}{10}

\bibitem{Najm17}
E.~Najm, R.~D. Yates, and E.~Soljanin.
\newblock Status updates through {M/G/1/1} queues with {HARQ}.
\newblock In {\em IEEE ISIT}, June 2017.

\bibitem{Soysal18}
A.~Soysal and S.~Ulukus.
\newblock Age of information in {G/G/1/1} systems.
\newblock In {\em Asilomar Conference}, November 2019.

\bibitem{Yates17b}
R.~D. Yates, P.~Ciblat, A.~Yener, and M.~Wigger.
\newblock Age-optimal constrained cache updating.
\newblock In {\em IEEE ISIT}, June 2017.

\bibitem{Bastopcu20d}
M.~Bastopcu and S.~Ulukus.
\newblock Information freshness in cache updating systems.
\newblock {\em IEEE Transactions on Wireless Communications}.
\newblock Early access.

\bibitem{Farazi18}
S.~Farazi, A.~G. Klein, and D.~R. Brown~III.
\newblock Average age of information for status update systems with an energy
  harvesting server.
\newblock In {\em IEEE Infocom}, April 2018.

\bibitem{Wu18}
X.~Wu, J.~Yang, and J.~Wu.
\newblock Optimal status update for age of information minimization with an
  energy harvesting source.
\newblock {\em IEEE Transactions on Green Communications and Networking},
  2(1):193--204, March 2018.

\bibitem{Ayan19}
O.~Ayan, M.~Vilgelm, M.~Kl\"{u}gel, S.~Hirche, and W.~Kellerer.
\newblock Age-of-information vs. value-of-information scheduling for cellular
  networked control systems.
\newblock In {\em ACM ICCPS}, April 2019.

\bibitem{Baknina18b}
A.~Baknina, O.~Ozel, J.~Yang, S.~Ulukus, and A.~Yener.
\newblock Sending information through status updates.
\newblock In {\em IEEE ISIT}, June 2018.

\bibitem{Leng19}
S.~Leng and A.~Yener.
\newblock Age of information minimization for an energy harvesting cognitive
  radio.
\newblock {\em IEEE Transactions on Cognitive Communications and Networking},
  5(2):427--439, June 2019.

\bibitem{Arafa19c}
A.~Arafa and S.~Ulukus.
\newblock Timely updates in energy harvesting two-hop networks: Offline and
  online policies.
\newblock {\em IEEE Transactions on Wireless Communications}, 18(8):4017--4030,
  August 2019.

\bibitem{Gu20}
Y.~Gu, Q.~Wang, H.~Chen, Y.~Li, and B.~Vucetic.
\newblock Optimizing information freshness in two-hop status update systems
  under a resource constraint.
\newblock July 2020.
\newblock Available on arXiv: 2007.02531.

\bibitem{Arafa20a}
A.~{Arafa}, J.~{Yang}, S.~{Ulukus}, and H.~V. {Poor}.
\newblock Age-minimal transmission for energy harvesting sensors with finite
  batteries: Online policies.
\newblock {\em IEEE Transactions on Information Theory}, 66(1):534--556,
  January 2020.

\bibitem{Elmagid19}
M.~A. {Abd-Elmagid}, H.~S. {Dhillon}, and N.~{Pappas}.
\newblock A reinforcement learning framework for optimizing age of information
  in {RF}-powered communication systems.
\newblock {\em IEEE Transactions on Communications}, 68(8):4747--4760, May
  2020.

\bibitem{bastopcu_soft_updates_journal}
M.~Bastopcu and S.~Ulukus.
\newblock Minimizing age of information with soft updates.
\newblock {\em Journal of Communications and Networks}, 21(3):233--243, June
  2019.

\bibitem{Bastopcu20e}
M.~Bastopcu and S.~Ulukus.
\newblock Timely group updating.
\newblock November 2020.
\newblock Available on arXiv:2011.15114.

\bibitem{Ceran18}
E.~T. Ceran, D.~Gunduz, and A.~Gyorgy.
\newblock A reinforcement learning approach to age of information in multi-user
  networks.
\newblock In {\em IEEE PIMRC}, September 2018.

\bibitem{Yates17a}
R.~D. Yates and S.~K. Kaul.
\newblock The age of information: Real-time status updating by multiple
  sources.
\newblock {\em IEEE Transactions on Information Theory}, 65(3):1807--1827,
  March 2019.

\bibitem{Kadota18a}
I.~Kadota, A.~Sinha, E.~Uysal-Biyikoglu, R.~Singh, and E.~Modiano.
\newblock Scheduling policies for minimizing age of information in broadcast
  wireless networks.
\newblock {\em IEEE/ACM Transactions on Networking}, 26(6):2637--2650, December
  2018.

\bibitem{Hsu18b}
Y.~Hsu.
\newblock Age of information: {Whittle} index for scheduling stochastic
  arrivals.
\newblock In {\em IEEE ISIT}, June 2018.

\bibitem{Buyukates19b}
B.~Buyukates, A.~Soysal, and S.~Ulukus.
\newblock Age of information scaling in large networks with hierarchical
  cooperation.
\newblock In {\em IEEE Globecom}, December 2019.

\bibitem{Buyukates18b}
B.~Buyukates, A.~Soysal, and S.~Ulukus.
\newblock Age of information in multihop multicast networks.
\newblock {\em Journal of Communications and Networks}, 21(3):256--267, July
  2019.

\bibitem{Wang19a}
M.~Wang, W.~Chen, and A.~Ephremides.
\newblock Reconstruction of counting process in real-time: The freshness of
  information through queues.
\newblock In {\em IEEE ICC}, July 2019.

\bibitem{Bastopcu20a}
M.~Bastopcu and S.~Ulukus.
\newblock Who should {Google} {Scholar} update more often?
\newblock In {\em IEEE Infocom}, July 2020.

\bibitem{Sun17b}
Y.~Sun, Y.~Polyanskiy, and E.~Uysal-Biyikoglu.
\newblock Remote estimation of the {Wiener} process over a channel with random
  delay.
\newblock In {\em IEEE ISIT}, June 2017.

\bibitem{Yun18}
J.~{Yun}, C.~{Joo}, and A.~{Eryilmaz}.
\newblock Optimal real-time monitoring of an information source under
  communication costs.
\newblock In {\em IEEE CDC}, December 2018.

\bibitem{Kam20}
C.~{Kam}, S.~{Kompella}, and A.~{Ephremides}.
\newblock Age of incorrect information for remote estimation of a binary
  {M}arkov source.
\newblock In {\em IEEE Infocom}, July 2020.

\bibitem{Chakravorty18}
J.~{Chakravorty} and A.~{Mahajan}.
\newblock Remote estimation over a packet-drop channel with {Markovian} state.
\newblock {\em IEEE Transactions on Automatic Control}, 65(5):2016--2031, July
  2020.

\bibitem{Mayekar20}
P.~{Mayekar}, P.~{Parag}, and H.~{Tyagi}.
\newblock Optimal source codes for timely updates.
\newblock {\em IEEE Transactions on Information Theory}, 66(6):3714--3731,
  March 2020.

\bibitem{MelihBatu4}
M.~Bastopcu, B.~Buyukates, and S.~Ulukus.
\newblock Selective encoding policies for maximizing information freshness.
\newblock April 2020.
\newblock Available on arXiv:2004.06091.

\bibitem{Ramirez19}
D.~{Ramirez}, E.~{Erkip}, and H.~V. {Poor}.
\newblock Age of information with finite horizon and partial updates.
\newblock In {\em IEEE ICASSP}, May 2020.

\bibitem{Buyukates19c}
B.~{Buyukates} and S.~{Ulukus}.
\newblock Timely distributed computation with stragglers.
\newblock {\em IEEE Transactions on Communications}, 68(9):5273--5282,
  September 2020.

\bibitem{Zou19a}
P.~Zou, O.~Ozel, and S.~Subramaniam.
\newblock Optimizing information freshness through computation-transmission
  tradeoff and queue management in edge computing.
\newblock December 2019.
\newblock Available on arXiv: 1912.02692.

\bibitem{Bastopcu19}
M.~Bastopcu and S.~Ulukus.
\newblock Age of information for updates with distortion.
\newblock In {\em IEEE ITW}, August 2019.

\bibitem{Rajaraman18}
N.~Rajaraman, R.~Vaze, and R.~Goonwanth.
\newblock Not just age but age and quality of information.
\newblock December 2018.
\newblock Available on arXiv:1812.08617.

\bibitem{Bedewy19}
A.~M. Bedewy, Y.~Sun, S.~Kompella, and N.~B. Shroff.
\newblock Age-optimal sampling and transmission scheduling in multi-source
  systems.
\newblock In {\em ACM MobiHoc}, July 2019.

\bibitem{Banerjee20}
S.~Banerjee, R.~Bhattacharjee, and A.~Sinha.
\newblock Fundamental limits of age-of-information in stationary and
  non-stationary environments.
\newblock In {\em IEEE ISIT}, June 2020.

\bibitem{Yates14}
R.~D. Yates and D.~J. Goodman.
\newblock {\em Probability and Stochastic Processes}.
\newblock Wiley, 2014.

\bibitem{Boyd04}
S.~P. Boyd and L.~Vandenberghe.
\newblock {\em Convex Optimization}.
\newblock Cambridge University Press, 2004.

\end{thebibliography}
\end{document}